\documentclass[aps,prb,twocolumn,reprint,groupedaddress,showpacs]{revtex4-1}
\usepackage{graphicx}
\usepackage{verbatim}
\usepackage{amssymb,amsmath}

\newcommand{\ket}[1]{| #1\rangle}
\newcommand{\bra}[1]{\langle #1 |}

\newcommand{\abs}[1]{| #1 |}
\newcommand{\ex}[1]{\langle #1 \rangle}

\begin{document}


\title{Probability distributions of continuous measurement results for conditioned quantum evolution}


\author{A. Franquet}
\email[]{A.FranquetGonzalez@tudelft.nl}

\author{Yuli V. Nazarov}
\affiliation{Kavli Institute of Nanoscience, Delft University of Technology, 2628 CJ Delft, The Netherlands}


\date{\today}

\begin{abstract}
We address the statistics of continuous weak linear measurement on a few-state quantum system that is subject to a conditioned quantum evolution. For a conditioned evolution, both the initial and final states of the system are fixed: the latter is achieved by the post-selection in the end of the evolution. The statistics may drastically differ from  the non-conditioned case, and the interference between initial and final states can be observed in the probability distributions of measurement outcomes as well as in the average values exceeding the conventional range of non-conditioned averages. We develop a proper formalism to compute the distributions of measurement outcomes, evaluate and discuss the distributions in experimentally relevant setups.  

We demonstrate the manifestations of the interference between initial and final states in various regimes. We consider analytically simple examples of non-trivial probability distributions. We reveal peaks (or dips) at {\it half-quantized} values of the measurement outputs. We discuss in detail the case of zero overlap between initial and final states demonstrating anomalously big average outputs and {\it sudden jump} in time-integrated output. 

We present and discuss the numerical evaluation of the probability distribution aiming at extending the analytical results and  describing a realistic experimental situation of a qubit in the regime of resonant fluorescence.

\end{abstract}

\maketitle

\section{Introduction}
\label{sec:intro}
The concept of measurement is one of the most important, characteristic, and controversial parts of quantum mechanics. Due to the intrinsically probabilistic nature of the measurement and associated paradoxes,\cite{Leggett} it continues to attract research attention and stimulate new experiments. The ability to control a quantum system that is of increasing importance in the context of quantum information processing, requires an adequate yet sufficiently general description of the measurement process. Such description is provided by the theory of continuous weak linear measurement (CWLM), where a sufficiently weak coupling between the quantum system and multiple degrees of freedom of a detector mediates their entanglement and results in conversion of discrete quantum information into continuous time-dependent readings of the detector.\cite{CWLM0,CWLM1,CWLM2,CWLM25,NazWei,CWLM3,CWLM4}
The description  follows from the general linear response theory and gives an explicit connection between quantum measurement and quantum noise.\cite{QNoise}\\
Recent experimental advances have made possible the efficient continuous measurement and monitoring of elementary quantum systems (qubits) giving the information on individual quantum trajectories.\cite{Devoret, SiddiqiSingle, SiddiqiEntanglement}The individual traces of quantum evolution can be post-selected by a projective measurement at the end of evolution, thus enabling the experimental investigation of conditioned quantum evolution where both initial and final states are known.\cite{Huard, DiCarlo, SiddiqiMapping, SiddiqiMolmer}\\

For experimentally relevant illustrations, we concentrate in this paper on a setup of resonance fluorescence.\cite{Huard} In this setup, a transmon qubit with ground state $\ket{g}$ and excited state $\ket{e}$ is enclosed in a non-resonant three-dimensional (3D) superconducting cavity connected to two transmission lines. A resonant field drives the qubit via the weakly coupled line, while most of the fluorescence signal exits via the other line which is coupled strongly. The amplitude of the signal is proportional to $\sigma_-$, the average of the lowering operator $\hat{\sigma}_-=\ket{g}\bra{e}$ of the qubit, and oscillates with the Rabi frequency $\Omega$ set by the resonant drive.\\
A heterodyne detection setup is used to measure this signal. The measurement proceeds in many runs of equal time duration. At each run,
the qubit is prepared in a state $\ket{e}$ or $\ket{g}$ and the signal is monitored at the time interval $0<t<{\cal T}$. At the end of the interval, $t={\cal T}$, one can projectively measure the qubit to find it either  in the state $\ket{e}$ or $\ket{g}$ with high fidelity using a microwave tone at the bare cavity frequency.
With such a setup, the fluorescence signal can be interpreted as a result of a weak continuous measurement, that can be conditioned not only on an initial state but also on a final state by post-selecting with the result of the projective measurement. The authors have concentrated on the conditioned signal at a given moment of time that is averaged over many runs.  Its time traces reveal interference patterns interpreted in terms of weak values \cite{WeakValues} and associated with the interference of initial and final quantum states in this context.\cite{WisemanWeakValues, paststates}

The concept of weak values has been introduced in \cite{WeakValues} to describe the average result of a weak measurement subject to post-selection in a simplified setup. The authors have shown that the average measurement results may be paradoxically large as compared to the outputs of corresponding projective measurements. Since that, the concept has been extended in various directions, e.g. to account for the intermediate measurement strength, the Hamiltonian evolution of the quantum states during the measurement, see \cite{WeakReview1,WeakReview2} for review. In \cite{WisemanWeakValues}, the average measurement outputs have been investigated in the context of continuous weak measurement, this has been further elaborated in \cite{ContWeak1,ContWeak2,ContWeak3}. As to the detailed statistics of the measurement outcomes, in this context it has been considered only for simplified meter setups that correspond to measuring the light intensities in quantum optics.\cite{WeakReview1,WeakReview2}
There is a tendency to term "weak value" a result of any weak measurement that involves post-selection. This may be confusing in general. For instance, the duration of a weak measurement can exceed the relaxation time of the system measured. The averaged measurement output in this case is not affected by post-selection and equals to the expectation value of the operator measured with the equilibrium density matrix. This is very far from the original definition of weak values\cite{WeakValues}. We prefer to stick to the original definition.

We notice that the experiment discussed gives access not only to the conditioned averages, but also to the conditioned statistics of the measurement results. For instance, at each run one can accumulate the output signal on a time interval that is  $(0,{\cal T})$ or a part of it and record the results. After many runs, one makes a histogram of the records that depends on the initial as well as on the final state of the qubit.

This article elaborates on the method to evaluate the distribution of the accumulated signal and gives the detailed theoretical predictions of the conditioned statistics for examples close to the actual experimental situation, and in a wide range of parameters.

In this Article, we put forward and investigate two signatures of the conditioned statistics. First is the {\it half-quantized} measurement values. A non-conditioned CWLM distribution under favorable circumstances peaks at the values corresponding to quantized values of the measured operator, in full correspondence with a text-book projective measurement. We demonstrate that a conditioned distribution function displays peculiarities --- that are either peaks or dips --- at {\it half-sums} of the quantized values.

Second signature pertains the case of zero or small overlap between initial and final state and time intervals that are so short as the wave function of the system does not significantly change. In this case, we reveal unexpectedly large values of the cumulants of the distribution function of time-integrated outputs for such short intervals, that we term {\it sudden jump}. For the average value of the output, the fact that it may by far exceed the values of typical outcome of a projective measurement, can be understood from the weak value theory \cite{WeakValues}. We extend these results to the distributions of the output and reveal the role of decoherence at small time intervals.

We stress that the signatures by itself present no new phenomenon. Rather, the basic quantum phenomena like interference manifest themselves in these signatures in the context of CWLM statistics. As such, we permit a re-interpretation of these phenomena in the context considered.

Our approach to the CWLM statistics is based on the theory of full counting statistics in the extended Keldysh formalism.\cite{QTBook} The statistics of measurements of $\int dt \hat{V}(t)$, $V(t)$ being a quantum mechanical variable representing linear degrees of freedom of the environment, are generated via a characteristic function method and the use of counting field technique. It provides the required description of the whole system consisting of the measured system, the environment and detectors.\\

Here we develop this formalism first introduced in,\cite{NazKin, NazWei} to include the conditioned evolution. We focus on the pre- and post-selected measurements. In this case, a quantum system  is initially prepared in a specific state. After that, it is subject of CWML during a time interval ${\cal T}$. The post-selection  in a specific state takes place in the end of the procedure. We show that the evolution of a qubit whose past and future states are known can be inferred and understood from the measured statistics of measurement outcomes. The measurement of the statistics can reveal purely quantum features in experimentally relevant regimes.\\

 We show how interference arises even at relatively small time scales and how the information about the initial qubit state is lost during the time evolution making the interference to vanish at sufficiently long time scales. We exemplify how different features in the distributions can be understood as the manifestations of the qubit evolution during the measurement. And we numerically study various parameter regimes of interest in the case of a measurement of a single observable.
 
 Actually, we show with our results that one can have very detailed theoretical predictions of CWLM distributions that can account for every detail of the experiment. This enables investigation and characterization of quantum effects even if the choice of parameters is far from the optimal one and these effects are small.

The structure of the article is as follows. We develop the necessary formalism in Section \ref{sec:method}, starting from a Bloch-master equation for the qubit evolution that is augmented with counting fields to describe the detector statistics, and explain how the post-selection is introduced in this scheme. The scheme can be applied to various experimental scenarios, in particular we focus on the setup described in \cite{Huard}. It is important to illustrate how the Cauchy-Schwartz inequalities  impose restrictions on the parameters entering the Bloch-master equations, this resulting in several different time scales. 
In Section \ref{sec:simple} we examine a measurement of a general observable and explain how the half-quantized peculiarities arise in the distributions of measurement outcomes depending on the initial and final state. In Section \ref{sec:suddenjump} we concentrate on the case of zero overlap and take the Hamiltonian dynamics into account to arrive at essentially non-Gaussian probability distributions. In these Sections, we mostly concentrate on a simple limit where the time interval ${\cal T}$ is much smaller than the typical  time scales of qubit evolution, this gives the opportunity for analytical results.
Next, we extend our study to longer time intervals. In the Section \ref{sec:decscale} we present numerical simulations at the scale of decoherence time for three relevant cases: the case of an ideal detector, and the  experimentally relevant case with and without detuning.
In Section \ref{sec:shortscale}, we concentrate on the time scales of Hamiltonian dynamics and experimentally relevant parameters. We conclude in the Section \ref{sec:conclusion}

\section{Method}
\label{sec:method}

The description of CWLM can be achieved by several methods, all of them taking into account the stochastic nature of the measurement process. In simplest situations like non-demolition measurements \cite{CWLM1} one can use the quantum filtering equation \cite{Belavkin}. More sophisticated approaches include effective action method \cite{CWLM0,CWLM4}, path integral formulation\cite{NazWei,CWLM3}, past states formalism \cite{paststates}. A powerful numerical method of experimental significance is the stochastic update equation \cite{trajectory} that allows to monitor density matrix taking into account the measurement results. In this method, the distribution of outcomes is obtained numerically by collecting statistics of  the realizations of "quantum trajectories". In contrast to this, the method of\cite{NazWei} permits the direct computation of the generating function of the probability distribution. 

The present goal is to formulate a method to compute probability distributions of a continuous measurement in the course of a conditioned quantum evolution. We will extend the method presented in \citep{NazWei} where the central object is a Bloch-master equation for the evolution of the measured quantum system that is augmented with the counting fields. 
Evaluating the trace of the extended density matrix from this equation as a function of the counting fields provides the generating function for the probability distribution of the detector output. To outline the formalism,  we will focus first on the simplest setup where a single detector measures a single qubit variable $\hat{{\cal O}}$. In the end of the section we will give a generalization to the case of two variables.\\

In general, the dynamics of an isolated quantum system are governed by a Hamiltonian $\hat{H}_{q}$. For a realistic system, weak interaction with an environment representing the outside world will generate decoherence and relaxation . In the CWLM paradigm, the quantum system is embedded in a linear environment described in the same manner by a Hamiltonian $\hat{H}_{d}$. The quantum system interacts with the environment via a coupling Hamiltonian $\hat{H}_{c}$, 

\begin{equation}
\label{eq1}
\hat{H} = \hat{H}_{q} + \hat{H}_{c} + \hat{H}_{d}
\end{equation}
with

\begin{equation}
\label{eq2}
\hat{H}_{c} = \hat{{\cal O}}\hat{Q},
\end{equation}

$\hat{{\cal O}}$ being an operator in the space of the quantum system, that value is to be measured.
Since $\hat{H}_{d}$ is a Hamiltonian of a linear system, it can generally be represented by a boson bath Hamiltonian. The input of the detector is characterized by an {\it input} variable $\hat{Q}$ that is linear in boson fields. The output of the detector is represented by the {\it output} variable $\hat{V}$ that is also linear in boson fields.\\
The dynamics and statistics of the measurement process are fully characterized by the two-time correlators of the operators $\hat{Q}(t)$, $\hat{V}(t)$. If we assume the qubit dynamics is slower than a typical time scale of the environment, the four relevant quantities correspond to zero-frequency values of the correlators,
\begin{subequations}
\label{corr}
\begin{eqnarray}
S_{QQ} =& \frac{1}{2}{\displaystyle \int_{-\infty}^{t}}dt'\left\langle\left\langle\hat{Q}(t)\hat{Q}(t')+\hat{Q}(t')\hat{Q}(t)\right\rangle\right\rangle,\\*
S_{QV} =& \frac{1}{2}{\displaystyle \int_{-\infty}^{t}}dt'\left\langle\left\langle\hat{Q}(t)\hat{V}(t')+\hat{V}(t')\hat{Q}(t)\right\rangle\right\rangle,\\*
S_{VV} =& \frac{1}{2}{\displaystyle \int_{-\infty}^{t}}dt'\left\langle\left\langle\hat{V}(t)\hat{V}(t')+\hat{V}(t')\hat{V}(t)\right\rangle\right\rangle,\\*
a_{VQ} =& -\frac{i}{\hbar}{\displaystyle \int_{-\infty}^{t}}dt'\left\langle[\hat{V}(t),\hat{Q}(t')]\right\rangle,\\*
a_{QV} =& -\frac{i}{\hbar}{\displaystyle \int_{-\infty}^{t}}dt'\left\langle[\hat{Q}(t),\hat{V}(t')]\right\rangle.
\end{eqnarray}
\end{subequations}
where $\langle\langle\hat{A}\hat{B}\rangle\rangle = \langle(\hat{A}-\langle\hat{A}\rangle)\rangle\langle(\hat{B}-\langle\hat{B}\rangle)\rangle$ for any pair of operators $\hat{A},\hat{B}$.\\
These four quantities define  the essential characteristics of the measurement process and have the following physical meaning. $S_{QQ}$ is the noise of the input variable. It is responsible for the inevitable measurement back action and associated decoherence of the qubit. $S_{VV}$ is the output variable noise: it determines the time required to measure the detector outcome with a given accuracy. The cross noise $S_{QV}$ quantifies possible correlations of these two noises. The response function $a_{VQ}$ determines the detector gain: it is the susceptibility relating the detector output to the qubit  variable measured, $\ex{\hat{V}}=a_{VQ}\ex{\hat{{\cal O}}}$. The response function $a_{QV}$ is correspondingly the reverse gain of the detector: it gives the change of the qubit variable proportional to the detector reading. Conforming to the assumption of slow qubit dynamics, the noises are white and responses are instant.\\
The values of these noises and responses are restricted by a Cauchy-Schwartz inequality, \citep{QNoise}

\begin{equation}
\label{ineq}
S_{QQ}S_{VV}-\abs{S_{QV}}^{2}\geq\frac{\hbar^{2}}{4}\abs{a_{VQ}-a_{QV}}^{2}.
\end{equation}

For a simple system like a single qubit it is natural to make the measured operator dimensionless, with eigenvalues of the order of one, or, even better, $\pm 1$.
With this, one can define and relate the dephasing rate $2 \gamma = 2S_{QQ}/\hbar^{2}$ and the acquisition time $t_{a}\equiv 4 S_{VV}/\abs{a_{VQ}}^{2}$ required to measure the variable with  ${\cal O}$ with a relative accuracy $\simeq 1$. If one further assumes the direct gain to be much larger than the reverse gain, $a_{VQ}\gg a_{QV}$, it is implied by the central equation of \citep{QNoise}, Eq. \eqref{eq3}, 
\begin{equation}
\gamma t_{a}\geq 1
\end{equation}
This figure of merit shows that one cannot measure a quantum system without dephasing it.\\

The statistics of the detector variable $\hat{V}$ can be evaluated with introducing a counting field $\chi(t)$ coupled to the output variable $\hat{V}$. This field plays the role of the parameter in the generating function $C({\chi(t)})$ of the probability distribution of the detector readings $V(t)$.\\
This generating function is computed in the extended Keldysh scheme \cite{QTBook} where the evolution of the "ket" and "bra" wave functions is governed by different Hamiltonians, $\hat{H}^{+}$ and $\hat{H}^{-}$ respectively. The extra term describing interaction with the counting field reads $\hat{H}^{\pm} = \hat{H} \pm \hbar \chi(t)\hat{V}(t)/2$. The generating function has then the form
\begin{equation}
\label{gen1}
C(\{\chi(t)\}) = \text{Tr}_{q}\left(\hat{\rho}(\{\chi(t)\})\right),
\end{equation}
 $\hat{\rho}$ being a quasi-density matrix of the qubit in the end of evolution, 
\begin{equation}
\label{gen2}
\hat{\rho}(\chi; t) =  \text{Tr}_{d}\left(\overrightarrow{T}e^{-i/\hbar\int dt \hat{H}^{-}}\hat{\rho}(0)\overleftarrow{T}e^{+i/\hbar\int dt \hat{H}^{+}}\right).
\end{equation}
Here, $\text{Tr}_{q}(\cdots)$ and $\text{Tr}_{d}(\cdots)$  denote the trace over qubit and detector variables, respectively, and $\overrightarrow{T}$($\overleftarrow{T}$) denotes time (reversed) ordering in evolution exponents. $\hat{\rho}(0)$ is the initial density matrix for both qubit and detector systems.\\

Assuming white noises and instant responses, one can derive an evolution  Bloch-master equation for the quasi-density matrix that is local in time, like Eq. (13) in \citep{NazWei}. For the simplest setup, under assumption of a single coupling operator $\hat{{\cal O}}$ it reads:

\begin{eqnarray}
\label{eq3}
\frac{\partial\hat{\rho}}{\partial t} =& -\frac{i}{\hbar}[\hat{H}_{q},\hat{\rho}] - \frac{S_{QQ}}{\hbar^{2}}\mathcal{D}[\hat{\cal O}]\hat{\rho} -\frac{\chi^{2}(t)}{2}S_{VV}\hat{\rho} \\*
\nonumber
& -\frac{S_{QV}}{\hbar}\chi(t)[\hat{\rho},\hat{\cal O}] + \frac{i a_{VQ}\chi(t)}{2}[\hat{\rho},\hat{\cal O}]_{+}.
\end{eqnarray}
Here, $[,]$ and $[,]_{+}$ refer to commutator and anti-commutator operations respectively and $\mathcal{D}[\hat{A}]\hat{\rho}\equiv\left(\frac{1}{2}[\hat{A}^{\dagger}\hat{A},\hat{\rho}]_{+}-\hat{A}\hat{\rho}\hat{A}^{\dagger}\right)$. Here we have also assumed  $a_{VQ}\gg a_{QV}$, a general condition for a good amplifier. A single coupling operator is an idealization, in a more realistic situation, the quantum system is also coupled to the environment with other degrees of freedom not related to the equation, this is manifested as intrinsic relaxation and decoherence. This modifies the above equation. 

We give the concrete form of this equation for the experimental situation of \cite{Huard}.
There is a qubit with two levels split in $z$-direction under conditions of strong resonant drive that compensates the splitting of the qubit levels. The effective  Hamiltonian reads
\begin{equation}
\hat{H}_{q}=\frac{\hbar}{2}\Omega\hat{\sigma}_{x}+ \frac{\hbar}{2}\Delta\hat{\sigma}_{z},
\end{equation}
$\Omega$ being the Rabi frequency proportional to the amplitude of the resonant drive, and $\Delta$ being the detuning of the drive frequency from the qubit energy splitting. The interaction with the environment induces  decoherence, excitation and relaxation of the qubit, with the rates $\gamma_{d},\gamma_{\uparrow},\gamma_{\downarrow}$ respectively. The measured quantity is the amplitude of the irradiation emitted from the qubit, so ${\cal O}$ is convenient to choose to be either $\sigma_x$ or $\sigma_y$. 
With this, the equation reads
\begin{eqnarray}
\label{eq3exp}
\frac{\partial\hat{\rho}}{\partial t} =& -\frac{i}{\hbar}[\hat{H}_{q},\hat{\rho}] -\gamma_{d}\mathcal{D}[\hat{\sigma}_{z}]\hat{\rho}-\gamma_{\uparrow}\mathcal{D}[\hat{\sigma}_{+}]\hat{\rho}\\*
\nonumber
&-\gamma_{\downarrow}\mathcal{D}[\hat{\sigma}_{-}]\hat{\rho} -\frac{S_{QV}}{\hbar}\chi(t)[\hat{\rho},\hat{{\cal O}}]\\*
\nonumber
& + \frac{i a_{VQ}\chi(t)}{2}[\hat{\rho},\hat{{\cal O}}]_{+}-\frac{\chi^{2}(t)}{2}S_{VV}\hat{\rho},
\end{eqnarray}
$\hat{\sigma}_{+}$ ($\hat{\sigma}_{-}$) being the rising and lowering operators of the qubit, and $\hat{\sigma}_{z}=\ket{e}\bra{e}-\ket{g}\bra{g}$ the standard Pauli operator.\\
The rates and noises are restricted by the following Cauchy-Schwartz inequality: $\frac{1}{4}\left(\gamma_{\uparrow}+\gamma_{\downarrow}\right)S_{VV}-\abs{S_{QV}}^{2}\geq\frac{\hbar^{2}}{4}\abs{a_{VQ}}^{2}$.
All the parameters entering the equation can be characterized from experimental measurements.
 We provide an example of concrete  values in Section \ref{sec:decscale}.\\


We will concentrate on a single measurement during a time interval $(0,{\cal T})$. To define an output of such measurement, we accumulate the time-dependent detector output during this time interval and normalize it by the same interval, $V \equiv \frac{1}{{\cal T}}\int_{0}^{{\cal T}} V(t') dt'$. The counting field $\chi(t)$ corresponding to this output is conveniently constant , $\chi(t)\equiv \chi$ on the time interval and 0 otherwise. Our goal is to evaluate the probability distribution $P(V)$ of the measurement results, conditioned to an initial qubit state given by $\hat{\rho}(0)$, and to a post-selection of the qubit in a specific state $\ket{\Psi}$ at the time moment ${\cal T}$. This involves the projection on the state $\ket{\Psi}$, represented by the projection operator $\hat{P}_{\Psi}=\ket{\Psi}\bra{\Psi}$ .\\
The probability distribution of the detector outcomes with no regard for the final qubit state can be computed from the generating function defined by Eq. \eqref{gen1},

\begin{equation}
\label{eq5}
P(V) = \frac{{\cal T}}{2\pi}\int d\chi e^{-i\chi V{\cal T}}C(\chi; {\cal T}).
\end{equation}

The joint statistics are extracted from the quasi-density matrix $\hat{\rho}(\chi; {\cal T})$ at the end of the interval.\\ 
Upon the post-selection, the quasi-density matrix is projected on the final state measured, $\hat{P}_{\Psi}\hat{\rho}(\chi; {\cal T})$, so  the conditioned generating function of the detector outcomes reads as
\begin{equation}
\label{eq6}
\tilde{C}(\chi; {\cal T}) = \frac{\text{Tr}_{q}(\hat{P}_{\Psi}\hat{\rho}(\chi; {\cal T}))}{\text{Tr}_{q}(\hat{P}_{\Psi}\hat{\rho}(\chi=0; {\cal T}))}.
\end{equation}
where the proper normalization is included.\\
This is the second central equation in our method. Together with Eq. \eqref{eq3} it permits an efficient evaluation of the conditioned probability distributions as the Fourier transform of this generating function.\\

Sometimes it is convenient to normalize the time-integrated output introducing ${\cal O} = V/a_{VQ}$ that immediately corresponds to the eigenvalues of $\hat{\cal O}$ (We stress that ${\cal O})$ are coming from the averaging of an environmental operator rather than $\hat{\cal O}$.

In this Article, we will concentrate on the distributions of a single variable. For completeness, we mention that the approach can be extended to joint 
statistics of simultaneous measurement of two non-commuting observables, e.g. $\hat{\sigma}_{x}$ and $\hat{\sigma}_{y}$. For the case of identical but independent detectors with associated output variables $\hat{V}_{x}, \hat{V}_{y}$ and counting fields $\chi_{x}(t),\chi_{y}(t)$ the corresponding equation reads( $i$ labels $\lbrace x,y\rbrace$)
\begin{eqnarray}
\label{eq3_2}
&\frac{\partial\hat{\rho}}{\partial t} = -\frac{i}{\hbar}[\hat{H}_{q},\hat{\rho}]-\sum_{i}\frac{S_{QQ}(i)}{\hbar^{2}}\mathcal{D}[\hat{\sigma}_{i}]\hat{\rho}\\*
\nonumber
& -\sum_{i}\left(\frac{S_{QV}}{\hbar}\chi_{i}(t)[\hat{\rho},\hat{\sigma}_{i}] +\frac{i a_{VQ}\chi_{i}(t)}{2}[\hat{\rho},\hat{\sigma}_{i}]_{+} -\frac{\chi_{i}^{2}(t)}{2}S_{VV}\hat{\rho}\right).
\end{eqnarray}
\\
for  the situation where the qubit decoherence is due to the detector back actions only. 
The parameters are restricted by inequalities similar to Eq. \eqref{ineq} for each set of noise and response functions corresponding to a given detector.\\
The form of this equation that can account for the realistic experimental situation \citep{Huard} is similar to Eq. \eqref{eq3exp}:
\begin{eqnarray}
\label{eq3_2exp}
&\frac{\partial\hat{\rho}}{\partial t} = -\frac{i}{\hbar}[\hat{H}_{q},\hat{\rho}] -\gamma_{d}\mathcal{D}[\hat{\sigma}_{z}]\hat{\rho}-\gamma_{\uparrow}\mathcal{D}[\hat{\sigma}_{+}]\hat{\rho}\\*
\nonumber
& -\gamma_{\downarrow}\mathcal{D}[\hat{\sigma}_{-}]\hat{\rho}-\sum_{i}\frac{S^{(i)}_{QV}}{\hbar}\chi_{i}(t)[\hat{\rho},\hat{\sigma}_{i}]\\*
\nonumber
& +\sum_{i}\frac{i a_{VQ}^{(i)}\chi_{i}(t)}{2}[\hat{\rho},\hat{\sigma}_{i}]_{+} -\sum_{i}\frac{\chi_{i}^{2}(t)}{2}S^{(i)}_{VV}\hat{\rho},
\end{eqnarray}
where $i=x,y$ and we account for detector-dependent noises and response functions. Two inequalities put restrictions on the parameters involved:
\begin{subequations}
\label{ineq3}
\begin{eqnarray}
\frac{1}{4}\left(\gamma_{\uparrow}+\gamma_{\downarrow}\right)S^{(x)}_{VV}-\abs{{S_{QV}^{(x)}}}^{2}\geq&\frac{\hbar^{2}}{4}\abs{a_{VQ}^{(x)}}^{2},\\*
\frac{1}{4}\left(\gamma_{\uparrow}+\gamma_{\downarrow}\right)S_{VV}^{(y)}-\abs{S_{QV}^{(y)}}^{2}\geq&\frac{\hbar^{2}}{4}\abs{a_{VQ}^{(y)}}^{2}.
\end{eqnarray}
\end{subequations}

Here, we have assumed an ideal and fast post-selection so that the system measured is projected on a known pure state $|\Psi\rangle$. This is the case of the experimental setup \cite{Huard}. In reality, there can be errors in the post-selection. We note 
that such errors can also be accounted for in the formalism outlined. To this end, one replaces the projection operator $\hat{P}_{\Psi}$ with a density matrix-like Hermitian operator $\hat{\rho}_f$ satisfying ${\rm Tr } [\hat{\rho}_f]  = 1$. For instance, if after a faulty projection measurement with the result "1" the system is in a orthogonal state $|\Psi_2\rangle$ with probability $p_e$, the corresponding $\hat{\rho}_f$ reads
\begin{equation}
\hat{\rho}_f = (1-p_e) |\Psi_1\rangle\langle\Psi_1| + p_e |\Psi_2\rangle\langle\Psi_2| 
\end{equation}

\section{Half-quantization: a straightforward case}
\label{sec:simple}
The outcomes of an ideal projective measurement of a quantum variable $\hat{{\cal O}}$ are confined to the eigenvalues ${\cal O}_i$ of the corresponding operator. If a CWML approximates well this ideal situation, one expects the distribution of outcomes to peak near ${\cal O}_i$, and it is indeed so. In this Section, we argue that
if the measurement outcomes are conditioned on  a final state, the distribution also has peculiarities at {\it half-sums} $({\cal O}_i+{\cal O}_j)/2$ of the eigenvalues. 
We prove first this counter-intuitive statement for a restricting limiting case where  the measurement interval ${\cal T}$ is much smaller than the typical time scales of the system dynamics. The results are summarized in Eq. \ref{eq:half-sums}.
The resulting distributions may formally correspond to negative probabilities in the limit of vanishing overlap between initial and final state. To correct for this, and to extend the limits of validity to larger time intervals, we concentrate further on a specific but constructive case of non-demolition measurement. With this, we investigate the influence of decoherence on half-quantization. The results are given by Eq.  \ref{eq:results-nondemolition}.

To start, we take the measurement interval ${\cal T}$ to be much smaller than typical time scales of the quantum system dynamics. This immediately implies that the accuracy of the measurement will be too low to make it practically useful. However, the resulting distribution comes out of a straightforward calculation, since the state of the quantum system does not have time to change significantly during the measurement.

In Eq. \eqref{eq3} we may then neglect  all terms describing the dynamics and containing no $\chi(t)$  
Let us also assume no correlation between the noises of the input and output variables of the detector, $S_{QV} = 0$.\\
With this, Eq. \eqref{eq3} can be simplified to the following form 

\begin{equation}
\label{eq3simp}
\frac{\partial\hat{\rho}}{\partial t} = -\frac{\chi^{2}(t)}{2}S_{VV}\hat{\rho} + \frac{i a_{VQ}\chi(t)}{2}[\hat{\rho},\hat{{\cal O}}]_{+}.
\end{equation}
Let us concentrate on a piecewise-constant $\chi(t)\equiv \chi\Theta(t)\Theta({\cal T} - t)$ corresponding to the accumulation of the signal during the measurement interval.
We take $\hat{\rho}(\chi; 0)=\hat{\rho}(0)$ as the initial condition. 
After the time interval of the measurement ${\cal T}$, the quasi-density matrix becomes

\begin{equation}
\label{simp}
\hat{\rho}(\chi; {\cal T}) = e^{-\frac{S_{VV}}{2}\chi^{2}{\cal T}}e^{i\frac{a_{VQ}}{2}\chi{\cal T}\hat{{\cal O}}}\hat{\rho}(0)e^{i\frac{a_{VQ}}{2}\chi{\cal T}\hat{{\cal O}}}.
\end{equation}

The generating function of the outcome distribution is given by Eq. \eqref{eq6} and involves the projection $\hat{P}_{\Psi}$ on the final state $|\Psi\rangle$. The calculations are straightforward in the basis of the eigenstates of the operator $\hat{{\cal O}}$, $\hat{{\cal O}}|i\rangle = {\cal O}_i |i\rangle$. It is also convenient to normalize the output variable on the value of $\hat{{\cal O}}$ introducing a rescaled variable ${\cal O} \equiv V/a_{VQ}$. The resulting distribution is a linear superposition of shifted normal distributions 
\begin{equation}
g(x)=\frac{1}{\sigma \sqrt{2\pi}} \exp\left(-\frac{x^2}{2\sigma^2}\right)
\end{equation}
with the same variance $\sigma^2 =S_{VV}/({\cal T} a^2_{VQ})= t_a/4{\cal T}$,
\begin{eqnarray}
\tilde{P}({\cal O}) = \sum_i W_{ii} g({\cal O}-{\cal O}_i) + \nonumber \\
\sum_{i\ne j} W_{ij} g\left({\cal O}-\frac{{\cal O}_i+{\cal O}_j}{2}\right)
\label{eq:half-sums}
\end{eqnarray}
and the weights $W_{ij}$ given by
\begin{equation}
W_{ij} = \frac{\Psi_j \Psi^*_i \rho^{(0)}_{ij}}{\langle \Psi|\rho^{(0)}|\Psi\rangle}; \; \sum_{i,j}W_{ij}=1.
\end{equation}

Let us discuss this result. The terms of the first group are normal distributions centered at the eigenvalues of ${\cal O}_i$. The coefficients in front of these terms are proportional to the product of the initial probability to be in the state $i$, $\rho(0)_{ii}$, and the probability to be found in final state after being in the state $i$, $|\Psi_i|^2$. If there would be no quantum mechanics, the system on its way from initial to final state should definitely pass one of the eigenstates of $\hat{{\cal O}}$ shifting the measurement output by the corresponding eigenvalue.
The sum of the probabilities $W_{ii}$ would be $1$. In fact, it is not $1$: owing to quantum interference, the system does not have to pass a definite state $i$. One can say that "bras" and "kets" may pass the different states, and this shifts the output by a half-sum of the corresponding eigenvalues. These interference contributions disappear if there is no post-selection in the final state. Indeed, summing $W_{ij}$ over a complete basis of possible final states $|\Psi\rangle$ gives zero. These coefficients also disappear in case of diagonal $\hat{\rho}^{(0)}$
Although the form (\ref{eq:half-sums}) suggests that real values $W_{i,j}+W_{j,i}$ could be interpreted as "probabilities" of "half-quantized" outcomes, this does not work since these values can be negative as well as positive, and the contributions centered at half-quantized values can be peaks as well as dips. This is typical for an interference effect. The double peak structure of the distribution  has been discussed earlier in the context of CWLM \cite{CWLM1,CWLM2,CWLM26,NazWei} The interpretation in terms of half-quantization is an innovation of the present article. 

A double-peak probability distribution has been predicted in the context of post-selected measurements \cite{Inverse0,Inverse1}. 
While this effect is also based on interference, it is clearly distinct from the half-quantization considered here since it is observed for an operator with continuous spectrum and in fact, in distinction from the effect described here, permits a classical interpretation\cite{Inverse1}. The half-quantization also does not bear any resemblance with the 3-box paradox \cite{completedescription} since the latter involves a third quantum state absent in our setup.

Nevertheless, the interference signatures can be revealed by a close inspection of the probability distribution of the outcomes of the conditioned measurement. We notice that the limit of small ${\cal T}$ we presently concentrate on is not favorable for such inspection since the peaks (or dips) are hardly separated, ${\cal O}_i \ll \sqrt{\sigma}$, so that $P({\cal O}) \approx g({\cal O})$, that is, hardly depends on the quantum system measured. To enhance the effect, one would increase ${\cal T}$. However, at sufficiently large ${\cal T}$ the quantum system would relax to equilibrium, this suppresses the interference effects. Numerical calculations presented in Sections \ref{sec:decscale} and \ref{sec:shortscale} show that the interference contributions become quite pronounced in the case of intermediate ${\cal T}$.

\begin{figure*}[!ht]
\centering
\includegraphics[width=1.6\columnwidth]{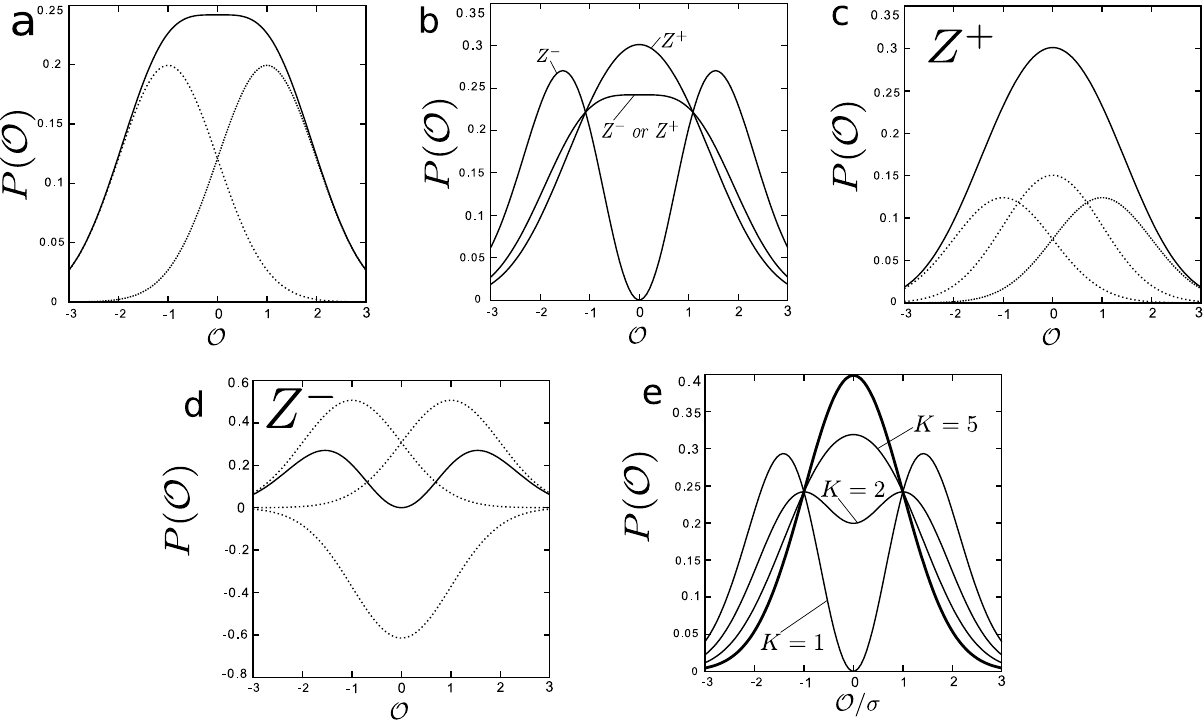}
\caption{Probability distributions of  CWLM outcomes of $\hat{\sigma}_{x}$ for relatively small duration ${\cal T} = 0.5 \Gamma_d$. The qubit is initialized in the $Z^{+}$ state. a. The distribution without post-selection consists of the two marginally separated Gaussian peaks shown by dotted lines. b. Conditioned distributions for $Z^+$ and $Z^-$ and the distribution without post-selection.  c.-d. Decomposition of the conditioned distributions into Gaussians (dotted curves). The Gaussians centered at $0$ manifest the half-quantization. e. The conditioned distribution for zero overlap (Eq. \ref{eq:zero-overlap}) in the limit of small ${\cal T}$ for different $K$. $K=1$ corresponds to ideal detector. }
\label{fig:simple-plots}
\end{figure*}

In this Section, we mention a special case where the interference effects become enhanced and significant even in the limit ${\cal T} \to 0$.   
This is the case of a small overlap between the initial state $\hat{\rho}(0)$ and the post-selected final state, $\ket{\Psi}$, $\langle \Psi |\hat{\rho}(0)|\Psi\rangle \to 0.$ \\
The coefficients $W_{ij}$ diverge upon approaching this limit, and Eq. \ref{eq:half-sums} becomes invalid giving a negative probability density.
To consider the case properly, we need to regularize Eq. \ref{simp} taking into account the dephasing which comes at least from the detector back-action. The simplest way to provide such regularization is to include dephasing produced by interaction with the same operator $\hat{{\cal O}}$. The resulting equation reads

\begin{equation}
\label{eq3simpS}
\frac{\partial\hat{\rho}}{\partial t} = -\frac{\chi^{2}(t)}{2}S_{VV}\hat{\rho} + \frac{i a_{VQ}\chi(t)}{2}[\hat{\rho},\hat{{\cal O}}]_{+} - \gamma\mathcal{D}[\hat{{\cal O} }]\hat{\rho}.
\end{equation}
It looks we have disregarded the Hamiltonian dynamics in Eq. \ref{eq3simpS}. This does not seem consistent since usually $H_q \gg \hbar \gamma$, this provides a common separation between the fast time-scales of Hamiltonian dynamics and longer time-scales of the decoherence and relaxation. We note that we do not have to disregard it in an important case of non-demolition measurement when $\hat{H}_q$ and $\hat{\cal O}$ commute. In this case, the only effect of the Hamiltonian dynamics is to provide time-dependent phase factors for non-diagonal elements of the density matrix. These trivial phase factors can be compensated by a proper rotation of the final state and the Hamiltonian dynamics can be gauged away from Eq. \ref{eq3simpS}. We address the relevant Hamiltonian dynamics in the next Section.

By virtue of the Cauchy-Schwartz inequality (\ref{ineq}), $\gamma \ge a^2_{QV}/4 S_{VV}$. Therefore it is convenient to characterize the dephasing rate $\gamma$ with dimensionless $K \equiv 4 \gamma S_{VV}/a^2_{QV} = \gamma t_a$, $K\geq1$, that characterizes the quality of the detector.

The equation is easily solved in the basis of eigenvalues of $\hat{{\cal O}}$. In comparison with Eq. \ref{simp}, each non-diagonal element $\rho_{ij}$ of the quasi-density matrix acquires an extra time-dependent suppression factor $\exp\left(-\gamma t \frac{({\cal O}_i - {\cal O}_j)^2}{2}\right)$. With this, the probability distribution is given by Eq. \ref{eq:half-sums}
with modified coefficients $W_{ij} \to \tilde{W}_{ij}$,
\begin{eqnarray}
\label{eq:results-nondemolition}
\tilde{W}_{ij} \equiv \frac{\Psi_j \Psi^*_i \rho_{ij} e^{-\gamma {\cal T}\frac{({\cal O}_i - {\cal O}_j)^2}{2}}}{\tilde{W}}; \\
\tilde{W} \equiv \sum_{i,j}\Psi_j \Psi^*_i \rho_{ij} e^{-\gamma {\cal T}\frac{({\cal O}_i - {\cal O}_j)^2}{2}} \nonumber
\end{eqnarray}

At any non-zero overlap, $P({\cal O}) \to g({\cal O})$ in the limit of ${\cal T} \to 0$. Let us concentrate on a special case of zero overlap, $\langle \Psi |\hat{\rho}^{(0)}|\Psi\rangle$=0, and let us note that  this also implies $\hat{\rho}^{(0)}|\Psi\rangle =0 $ by virture of positivity of the density matrix. In the limit of ${\cal T} \to 0$ the chance to find the system in the final state vanishes, 
$\tilde{W} \approx  \gamma {\cal T}\langle \Psi |{\hat{\cal O}}\hat{\rho}^{(0)}{\hat{\cal O}}|\Psi\rangle$. This divergency should be compensated by the terms $\propto {\cal T}$ that come from  expansion of $g({\cal O} - ({\cal O}_i +{\cal O}_j)/2)$ up to the second order in ${\cal O}_i$ as well as 
$\tilde{{W}}_{ij}$. The resulting distribution of the measurement outcomes for these rare events differs essentially from the normal one,
\begin{equation}
P({\cal O})= \left(1 +\frac{({\cal O}/\sigma)^2-1}{K} \right) g({\cal O}) \ne g({\cal O})
\label{eq:zero-overlap}
\end{equation}
For an ideal detector, $K=1$, the probability even vanishes at ${\cal O}=0$. For bigger decoherence exceeding the minimal one, $K\gg 1$, the interference term vanishes and $P({\cal O}) \approx g({\cal O})$.

We illustrate the content of this Section with some simple plots (Fig. \ref{fig:simple-plots}). We consider a qubit that is initially prepared in $Z^+$ state, $\hat{\sigma}_z|Z^+\rangle = |Z^+\rangle$. The measurement accesses the $x$-component of the qubit spin, ${\cal O}=\hat{\sigma}_x$.
After the measurement, the qubit is post-selected in either $Z^+$ or $Z^-$ state.
As it follows from the preceding discussion,
we expect the probability distribution of the outputs to be composed of the Gaussians centered at $\pm 1$, and also at the half sum of the eigenvalues, that is, at $0$. 

For the first four plots, we choose a relatively big ${\cal T} = 0.5 \gamma^{-1}$. Although this choice is  contrary to our assumptions, it  permits an easy visual resolution of the Gaussian peaks. We assume ideal detector $K=1$ and use Eq. \ref{eq3simpS} to evaluate the distributions. The distribution of the outcomes with no post-selection (Fig. \ref{fig:simple-plots}a.) is composed from two Gaussian peaks centered at $\pm 1$ that are hardly separated. 
The post-selected distributions differ much from each other and the original one (Fig. \ref{fig:simple-plots}b.) The distribution for $Z^-$ gives well-separated peaks while a single peak is seen in the distribution for $Z^{+}$. This is due to the negative or positive half-sum contribution as illustrated in Fig. \ref{fig:simple-plots}c. an d.

The Fig. \ref{fig:simple-plots}e. demonstrates the essential change of the conditioned distribution function for zero overlap. The distribution for ideal detector reaches zero, and approaches normal distribution upon increasing $K$.

To investigate in more detail the manifestations of the interference effects at longer time intervals $ \simeq t_a, \gamma^{-1}$ and in experimental conditions, in Section \ref{sec:decscale} we  numerically solve the evolution equations and compute the conditioned probability distributions. For this work, we concentrate on a single qubit.

\section{Sudden jump: a simple consideration}
\label{sec:suddenjump}

Let us now change the situation and consider the measurement of a variable that does not commute with the Hamiltonian. To simplify, we consider very small ${\cal T}$ such that the change of density matrix due to Hamiltonian dynamics is small. This is a more severe limitation than that used in the previous Section where ${\cal T}$ was only supposed to be smaller than the decoherence rate. Generally, this time interval is too small to measure anything and we expect the distribution to be close to $g({\cal O})$ thus to have a large spread. There is, however, an exceptional situation of zero overlap where after the measurement the state is projected on $|\Psi\rangle$ that is precisely orthogonal to the initial state $|i\rangle$, $\langle \Psi|i\rangle =0$. Let us concentrate on this situation and demonstrate a peculiarity of the output distribution which is best described as a {\it sudden jump} of the integrated output.

To give a clear picture, we first treat the situation completely disregarding the 
decoherence/relaxation terms, and take into account the Hamiltonian dynamics only. 
This seems relevant at such small ${\cal T}$. The general result is given by 
Eq. \ref{eq:suddengeneralresult} while a constructive case is given by \ref{eq:Cshort}.
This gives a sudden jump of cumulants while the attempt to derive the distribution 
results in a negative probability in an interval of outputs that increases with decreasing ${\cal T}$. To improve on this, we will sophisticate the treatment by including the decoherence. We reveal that the decoherence becomes important at very small time intervals  ${\cal T} \ll (\Omega^2 t_a)^{-1}$, that can be interpreted as a finite but small duration of the sudden jump. The resulting probability distribution is given by Eq. \ref{eq:distshorttimes} and is positive at any ${\cal T}$.

To start with, we disregard relaxation/decoherence terms in the evolution equation which seems relevant for such small ${\cal T}$ and owing to orthogonality, the projected $\rho(\chi)$ vanishes at ${\cal T} \to 0$ and is determined by the first-order corrections to bra- and ket wave functions,
\begin{equation}
\text{Tr}(\hat{P}_{\Psi}\hat{\rho}(\chi) )= \hbar^{-2}{\cal T}^2 \langle \Psi | \hat{H}^+_q | i\rangle \langle i |\hat{H}^-_q|\Psi \rangle e^{-\chi^2 {\cal T} S_{VV}/2}
\end{equation}
Here $H^{\pm} = H_q \pm \hbar \chi a_{VQ} \hat{\cal O}$.

The small factor ${\cal T}^2$ cancels upon normalization in Eq. \ref{eq6} so that the generating function of the conditioned output reads
\begin{equation}
\label{eq:suddengeneralresult}
\tilde{C}(\chi; {\cal T}) = \frac{\langle \Psi | \hat{H}^+_q | i\rangle \langle i |\hat{H}^-_q|\Psi \rangle}{|\langle \Psi | \hat{H}_q | i\rangle|^2} e^{-\chi^2 {\cal T} S_{VV}/2}
\end{equation}
We note that $\tilde{C}(\chi; {\cal T}\to 0) \ne 1$. Since the derivatives of $\ln \tilde{C}$ at $\chi \to 0$ are related to the cumulants $\kappa_n$ of the distribution of the integrated output 
$\int_0^{{\cal T}} dt \hat{V}(t)$. This implies that the cumulants of  the distribution of the integrated output do not vanish in the limit of short time interval: rather, there is a {\it sudden jump} of the integrated output not depending on the duration of the measurement. 
The jump occurs for the averaged output as well as for all cumulants. This is very counter-intuitive for a CWLM situation. In this case, one may expect that the integrated output in this limit is dominated
by the detector noise, so that $ \int_0^{{\cal T}} dt \hat{V}(t) \simeq {\cal T}^{1/2}$ ,
$\kappa_n \simeq {\cal T}^{n/2}$, and thus vanishes at ${\cal T} \to 0$.

To see this in more detail, let us turn to a concrete example. We consider a situation corresponding to \cite{Huard}: a qubit with the Hamiltonian $\hat{H}_{q}=\frac{\hbar}{2}\Omega\hat{\sigma}_{x}$. The initial and projected states 
are $Z^+$ and $Z^-$, respectively, and we measure the projection of the qubit on Y-axis, $\hat{{\cal O}} = \hat{\sigma}_y$.
In this case, 
\begin{equation}
\label{eq:Cshort}
\tilde{C}(\chi; {\cal T})= \left( 1-\frac{i\chi a_{VQ}}{\Omega}\right)^2 e^{-\chi^2 {\cal T} S_{VV}/2}
\end{equation}
In the limit ${\cal T} \to 0$ we obtain for the cumulants:
\begin{equation}
\kappa_n = \frac{\partial^{n}}{\partial (i\chi)^{n}} \ln \left( 1-\frac{i\chi a_{VQ}}{\Omega}\right)^2 = 2 (-1)^n\left(\frac{a_{VQ}}{\Omega}\right)^{n} (n-1)!
\end{equation}
We see a sudden jump in the cumulants of the time-integrated output.

The average value of the output ($\kappa_1$) is given by
\begin{equation} 
\label{eq:suddenjump}
a_{VQ}^{-1}\int_0^{{\cal T}} dt \langle \hat{V}(t)\rangle = -\frac{2}{\Omega};\;\bar{{\cal O}} = -\frac{2}{\Omega {\cal T}}.
\end{equation}
This corresponds to the time-averaged output $\propto {\cal T}^{-1}$ that can exceed by far the expected values of a projective measurement, $\pm 1$. 
Such anomalously big outputs are naturally associated with the weak values \cite{WeakValues}. Indeed, one can relate the above result with weak value conform to the definition \cite{WeakValues} if one takes into account the evolution of the quantum state during the measurement \cite{weakanddynamics}.  However, we need to stress that the full distribution of the outputs cannot be obtained with the traditional weak value formalism
and so far has not been obtained with its extensions \cite{ContWeak1,ContWeak2,ContWeak3} for continuous measurement. The method outlined here does not explicitly evoke the notion of weak values and provides a more elaborated description of a realistic measurement process.

 \begin{figure}[!ht]
\centering
\includegraphics[width=0.9\columnwidth]{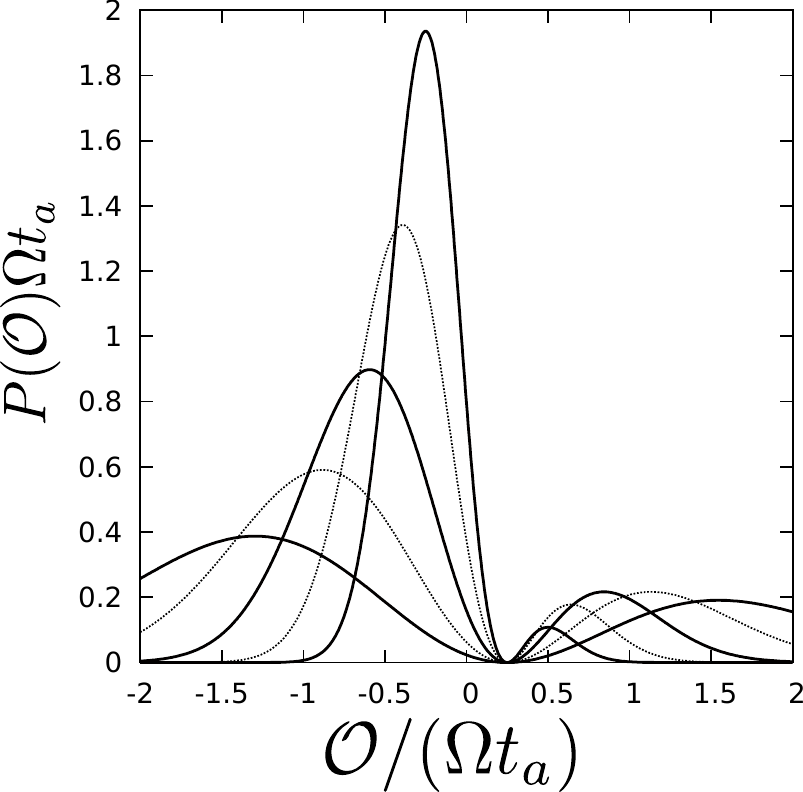}
\caption{Probability distributions of outputs (Eq. \ref{eq:distshorttimes}) in the sudden jump regime in case of an ideal detector. The alternating solid-dotted curves correspond to different ${\cal T} = (0.25,0.5,1.0,2.0,4.0) (\Omega^2 t_a)^{-1}$. Each curve consists of two peaks separated by a gap at ${\cal O}
=\Omega t_a/4$. The curves with bigger ${\cal T}$ are sharper, and the peaks become increasingly symmetric upon lowering ${\cal T}$.}\label{fig:shortscales}
\end{figure}

An attempt to derive from (\ref{eq:Cshort}) the  overall distribution of the time-averaged outputs yields
\begin{equation}
P({\cal O}) = \left(1+ \frac{\partial_{\cal O}}{\Omega {\cal T}}\right)^{2} g({\cal O})= \left(\left(1-\frac{{4 \cal O}}{\Omega t_a}\right)^2-\frac{4}{\Omega^2 {\cal T} t_a}\right)g({\cal O})
\end{equation}
 There is a problem with this expression: it is negative in an interval of ${\cal O}$, and at sufficiently small ${\cal T} \lesssim (\Omega^2 t_a)^{-1}$ this interval encompasses the body of the "distribution". This signals that the current approach must be corrected.  As we have seen in the previous Section, such correction most likely requires a proper account of the detector back-action that causes the decoherence of the qubit. 

It is unusual to expect a decisive role of decoherence at such small time scales. However, if we take into account the decoherence (second term in the r.h.s. of Eq. \ref{eq3} ), we obtain
\begin{equation}
\text{Tr}(\hat{P}_{\Psi}\hat{\rho}(\chi)) = \left(\gamma {\cal T}+ \frac{{\cal T}^2}{4} \left(\Omega - i a_{QV}\chi \right)^2\right) e^{-\chi^2 {\cal T} S_{VV}/2}
\end{equation}
Here, $\gamma \equiv S_{QQ}/\hbar^2$ is the corresponding decoherence rate. We see that the decoherence term may indeed compete with the term coming from Hamiltonian dynamics at short time intervals.  The Physical reason for this is that a decoherence term of this sort induces the relaxation in $Z$-basis. The relaxation brings the qubit to $Z^-$ faster than the Hamiltonian:  The probability to find the system in $Z^-$ is thus proportional to ${\cal T}$ in contrast to the probability $\propto {\cal T}^2$ induced by the Hamiltonian dynamics.

The resulting characteristic function reads
\begin{equation}
\tilde C(\chi)=\frac{ 4\gamma + {\cal T}\left(\Omega - i a_{QV}\chi \right)^2}{4\gamma+{\cal T}\Omega^2}e^{-\chi^2 {\cal T} S_{VV}/2}
\end{equation}
and gives the average output 
\begin{equation}
\label{eq:shorterscales}
\bar {{\cal O}}= - \frac{2 \Omega}{4\gamma + {\cal T}\Omega^2}
\end{equation}
The value of the average output thus saturates at $-\Omega/2\gamma \ll -1$ in the limit of small ${\cal T}\ll \gamma/\Omega^2$. So if the decoherence is taken into account, the change of the output averages is not really sudden. One can regard the small time scale $\gamma/\Omega^2$ of the saturation as a typical duration of the sudden jump of the time-integrated output. 

The probability distribution valid at all time scales $\ll \Omega^{-1}$ is given by
\begin{equation}
\label{eq:distshorttimes}
P({\cal O})=\frac{K-1 +({\cal T}/4t_a) \left(\Omega t_a - 4 {\cal O}\right)^2}{K+{\cal T}t_a \Omega^2/4} g({\cal O})
\end{equation}
 where we again introduce the dimensionless $K=\gamma t_a\geq 1$ that characterizes the quality of the detector. 
 The distribution is illustrated in Fig. \ref{fig:shortscales} for an ideal detector $K=1$ and various ${\cal T}$. In this case, the probability density is zero at ${\cal O}=\Omega t_a/4$.

If we compare the distributions (\ref{eq:zero-overlap}) and (\ref{eq:distshorttimes}), we see that the results of the previous Section are reproduced in the limit $\Omega \to 0$, as well as in the limit of ${\cal T} \ll (\Omega^2 t_a)^{-1}$ if we take $\sigma^2 = t_a/4 {\cal T}$. The distribution (\ref{eq:distshorttimes}) thus generalizes (\ref{eq:zero-overlap}) to the case where the Hamiltonian dynamics are relevant. 

To extend the results on larger time intervals $\simeq \Omega$  and on realistic conditions,  we  numerically solve the evolution equations in Section \ref{sec:shortscale} and compute the corresponding conditioned probability distributions. 

\section{Numerical results: long time scales}
\label{sec:decscale}
 
In Section \ref{sec:simple}, we have presented an analytical solution in the limit of small ${\cal T}$ and shown that it remains qualitatively valid for bigger ${\cal T}$, at least in the case of ideal detectors. We will extend these results evaluating the conditioned distributions numerically. We concentrate on longer measurement times where the qubit dynamics become important. We will take into account the effects of decoherence and relaxation, as well as the effects of strong qubit drive or detuning, all being important in experimental situations.\\

 In this Section, we address the distributions of the CWLM outcomes of a single variable at the time scales of the order of coherence/relaxation times and $t_a$. Generally, one can associate it with the qubit variable $\hat{{\cal O}}=\hat{\sigma}_{x}$. To start with, we assume zero detuning, that is, a qubit Hamiltonian of the form $\hat{H}_{q}=\frac{\hbar}{2}\Omega\hat{\sigma}_{x}$. In principle, we are now in the situation of a non-demolition measurement.


To start with, let us assume an idealized situation where all the decoherence is brought by the detector back action and its rate $\propto S_{QQ}$ assumes the minimum value permitted by the inequality (\eqref{ineq}). Since $\hat{H}_{q}=\frac{\hbar}{2}\Omega\hat{\sigma}_{x}$, the back-action does not interfere with free qubit dynamics causing transitions between the levels. In $\sigma_x$ representation, the diagonal elements of the density matrix remain unchanged keeping the initial probability to be in $X^{\pm}$ states while the non-diagonal ones oscillate with frequency $\Omega$ and decay with much slower rate $\Gamma_d \ll \Omega$. 

If we keep the final state fixed to $Z^{\pm}$, the interference contribution to the conditioned distributions will exhibit fast oscillations as function of ${\cal T}$ with a period $2\pi/\Omega$. It is proficient from both theoretical and experimental considerations to quench these rather trivial oscillations. We achieve this by projecting the qubit after the measurement on the states  $|\bar{Z}^{\pm}\rangle = e^{-i \hat{H}_q {\cal T}} |Z^{\pm} \rangle$ thereby correcting for the trivial qubit dynamics. In practice, such correction can be achieved by applying a short pulse rotating the qubit about $x$-axis right before the post-selection measurement. 
With this, the conditioned distribution of outcomes changes only at the time scale $t_a \simeq \Gamma^{-1}_d$, that is much longer than $\Omega^{-1}$, and the dynamics are described by Eq. \eqref{eq3simpS} with $\hat{\cal O}=\hat{\sigma}_x$. 

\begin{figure*}[t]
\centering
\includegraphics[scale=0.6]{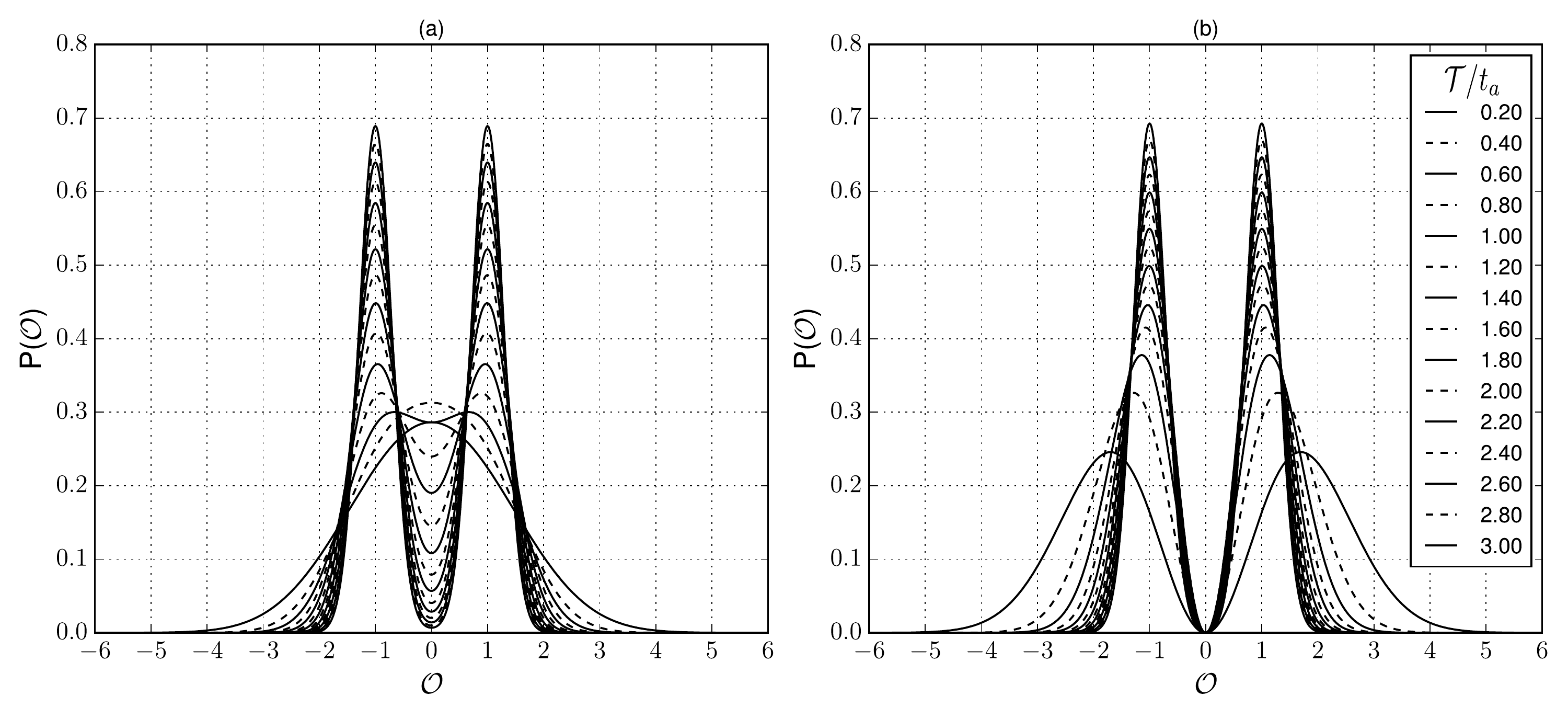}
\caption{Probability distributions of a $\hat{\sigma}_{x}$ CWLM  outcomes for the ideal measurement case for different ratios ${\cal T}/t_a$. The qubit is initially prepared in the $+Z$ state and, after ${\cal T}$, is post-selected either  in the $+Z$ state (a) or $-Z$ state (b).}
\label{fig2}
\end{figure*}

In Fig. \ref{fig2}, we give the plots of the probability distributions conditioned on $\bar{Z}^{\pm}$ for a series of measurement time intervals ${\cal T}$.
We see that 
 (different curves) are shown, for two cases in which the visibility of the interference feature is stronger, the case of equal preparation and post-selection, (a), and the case of orthogonal preparation and post-selection states, (b).\\
In this ideal situation, even for very small time intervals, the additional knowledge of the post-selection can lead to perfect resolution of the two eigenstates of the qubit variable (Fig. \ref{fig2} (b)). While for small time intervals the middle peak results in less resolution for the opposite choice of post-selected qubit state (Fig. \ref{fig2} (a)), at large time intervals, the detector back action has resulted in a complete decoherence of the qubit state and the interference signature disappears, making both distributions converge to two narrow peaks corresponding to either $+X$ or $-X$. This exemplifies how the knowledge of the qubit preparation is lost in time due to decoherence.

The fact that we see no difference between the distributions in this limit is a result of a symmetric choice we made with respect to the projections. Indeed, if we project on $\pm X$ instead, the distributions would consist of a single peak positioned at the value of ${\cal O}=\pm 1$. Generally, for projections on arbitrary pair of orthogonal superpositions of $X$ and $Z$, we expect in this limit different peak weights for two different projections. This difference, however, is of trivial origin and has nothing to do with the interference effects of interest. 
So we have made a symmetric choice to cancel it.

With this, the difference between the two distributions is due to interference only, that is, due to the half-quantized peak described in the previous Section. At smaller ${\cal T}$, the distributions take a very distinct shape: single-peak
for that conditioned on $+Z$, and double-peak for that conditioned on $-Z$. The half-quantization is dumped on the scale of the decoherence time, so the difference is seen only for ${\cal T} <t_a$.

The separation of the distribution onto two peaks in the limit of ${\cal T} \gg t_a$ is a signature of the ideal situation of a quantum non-demolition measurement where neither measurement nor any other agent induces the relaxation rates causing the transitions between the qubit states. In this situation, the density matrix efficiently relaxes to its equilibrium value $\rho_{eq}$at time interval ${\cal T}$, and the distribution of the  detector output tends to concentrate on the average value $\langle {\cal O}\rangle  = {\rm Tr}[\hat{\cal O} \hat{\rho}_{eq}]$
with decreasing width $\simeq \sqrt{t_a/{\cal T}}$.

Let us now turn to the analysis of the experimental situation. We use the general evolution equation Eq. \eqref{eq3exp} to compute the distributions and substitute the parameters $\gamma_{\downarrow} = (22.5 \mu\text{s})^{-1}, \gamma_{\uparrow} = (56 \mu\text{s})^{-1},\gamma_d = (15.6 \mu\text{s})^{-1}$ given in \cite{Huard}. The acquisition time comes from the measurement rate 
$2/t_{a}\approx(92 \mu\text{s})^{-1}$. This rate in fluorescence experiments can be characterized by two different methods both based on the estimation of the probability distribution for the integrated homodyne signal conditioned on the state of the qubit, see Appendix F in the supplementary material of \citep{QNoise}. The quality of the measurement setup is thus rather far from ideal, $K = t_a \gamma_d \approx 12$. Nevertheless we predict some measurable interference effects in the outcome distributions.

\begin{figure*}[!ht]
\centering
\includegraphics[scale=0.6]{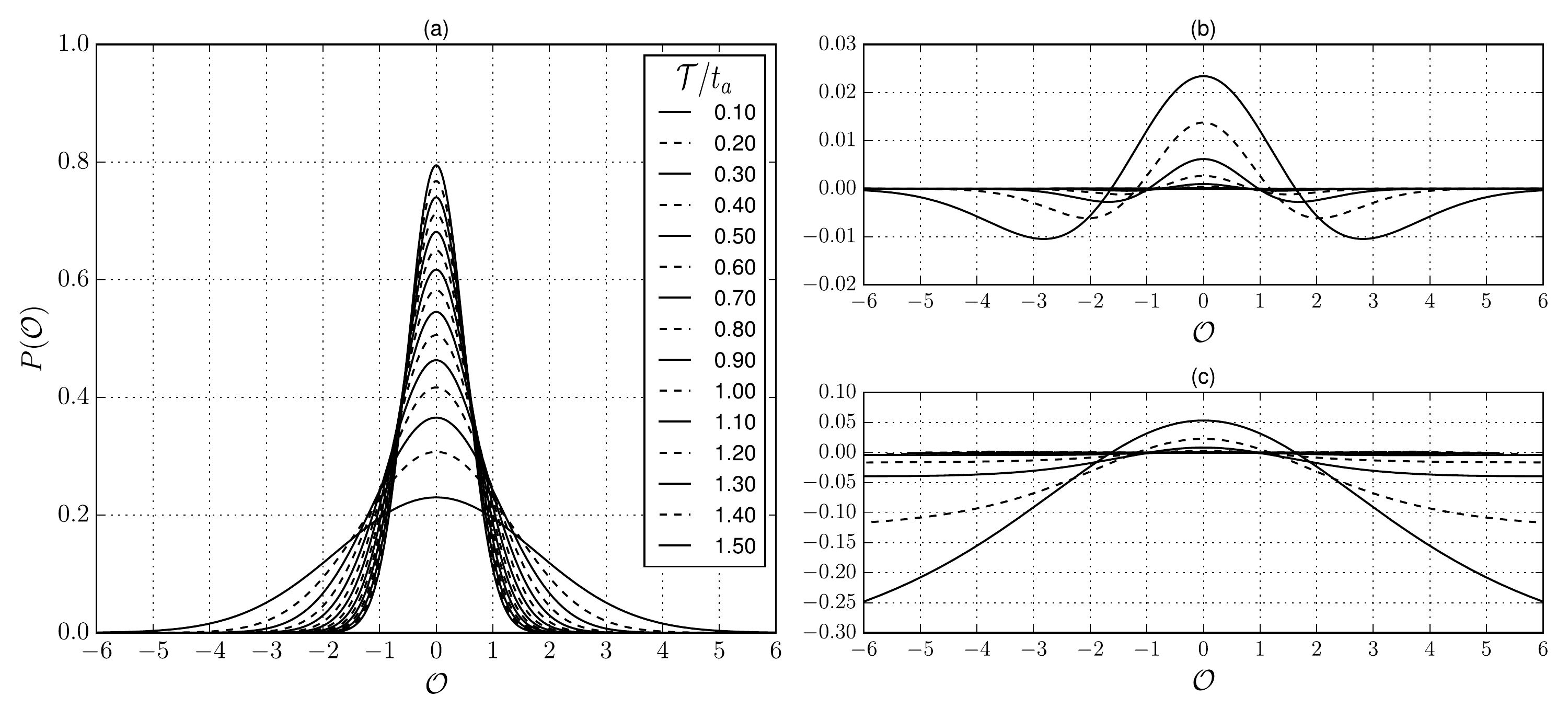}
\caption{Probability distributions of the output  $\hat{\sigma}_{x}$ CWLM  for the experimental setup of \cite{Huard} at various ${\cal T}$. Since the detection is far from ideal, the distributions conditioned on $\pm Z$ are not visually distinguishable, so we plot only one (a). However, the difference of the two distributions that is due to interference  (b) is sufficiently large to detect: the relative difference is about $10\%$  for small time intervals (top curve at ${\cal O}=0$ in (b))). In (c) we give the difference normalized to the sum of the probabilities. This quantity saturates at large ${\cal O}$.}
\label{fig3}
\end{figure*}

We plot in  Fig. \ref{fig3} the results for zero detuning.
There is no visible difference between the distributions, so in distinction from Fig. \ref{fig2}, we give only a single set of curves in Fig. \ref{fig3}. The curves for all ${\cal T}$ look dully Gaussian, no peak separation is visible. This is because of the low quality of the detector: the relaxation to the stationary density matrix $\hat{1}/2$ mainly takes place at a time interval shorter than the acquisition time, so most of the time the detector measures this featureless state. As to short ${\cal T}$, the distribution is too wide to manifest the 
features of the density matrix.

However, there are still observable signatures of interference. To reveal those, we plot in Fig. \ref{fig3} the difference of the probability densities for two projections. We see that at smallest ${\cal T} = 0.2t_a$ the relative difference achieves $0.1$ at ${\cal O}\approx 0$ and can be thus revealed from the statistics of several hundreds individual measurements. The shape of the difference suggests that the $P_-$ is pushed on both positive and negative values of ${\cal O}$ in comparison with $P_+$, in agreement with the previous findings.
The decoherence and relaxation quickly diminish the difference upon increasing ${\cal T}$. 

At big values of ${\cal O}$, the difference quickly decreases together with the distributions. In this respect, it is instructive to inspect the difference normalized on the sum of the probability densities, $C({\cal O})\equiv(P_+(\mathcal{O}) - P_-(\mathcal{O}))/(P_+(\mathcal{O}) + P_-(\mathcal{O}))$. This quantity gives the certainty with which one can distinguish two distributions from each other given a reading ${\cal O}$. The values $C=\pm 1$ would imply that the measurement is {\it certainly} post-selected with $\pm Z$. 
As we see from Fig. \ref{fig3} , the certainty saturates with increasing ${\cal O}$, reaches relatively large values at short ${\cal T}$, and fades away upon increasing ${\cal T}$. 

\begin{figure*}[!ht]
\centering
\includegraphics[scale=0.6]{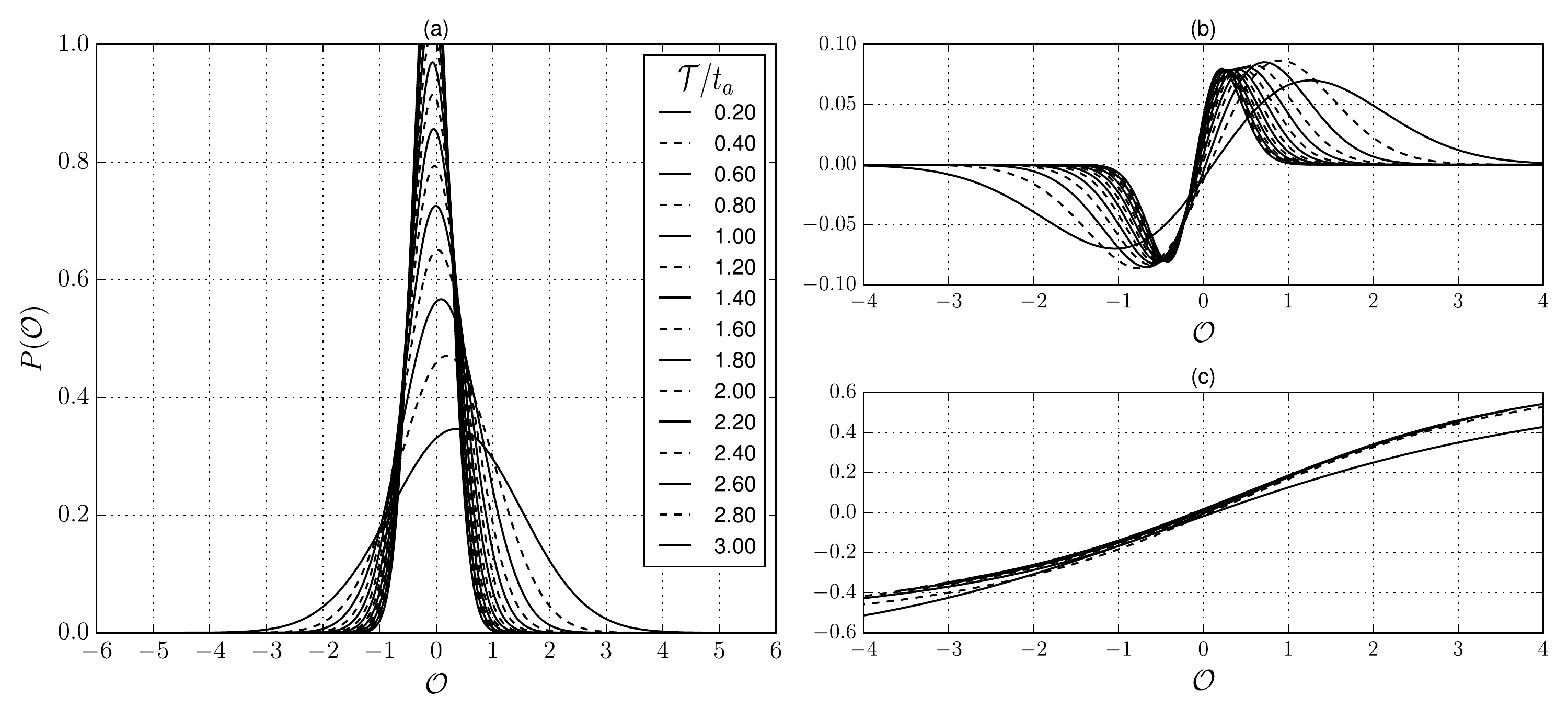}
\caption{Probability distributions of a $\hat{\sigma}_{x}$ weak measurement for experimental rates. Here a relatively large detuning $\Delta\approx1.7\Omega$ is introduced in the qubit Hamiltonian. The qubit is prepared in the $+Z$ state and post-selected, after a specific time interval given by each curve, in the $+Z$ state (a) or $-Z$ state (not in the Figure). The difference of this two probabilities appears to remain at rather big timescales while being remarkably large for small time intervals (wider curves in (b)) compared to the single distribution (a). Again, a good measure of this phenomena is the relative difference, here plotted in (c). Time in units of acquisition time $t_{a}$.}
\label{fig4}
\end{figure*}

Let us inspect the distributions at non-zero detuning. In this case, there is no reason to expect the ${\cal O} \to -{\cal O}$ symmetry in the distribution. We illustrate the situation in Fig. \ref{fig4} assuming relatively large detuning $\Delta = 1.7 \Omega$. This value is chosen to maximize $\langle{\cal O}\rangle$ for the equilibrium density matrix. In the plots of Fig. \ref{fig4}a,  we see a shift of the distribution maximum that tends to $\langle{\cal O}\rangle \approx -0.1$ at ${\cal T} \gg t_a$. The value of the 
shift does depend on ${\cal T}$ as well as on the post-selection state. 

If we concentrate on the difference of the probability distributions(Fig. \ref{fig4}b), we see the same order of magnitude as at zero detuning. However, the difference does not vanish in the limit of big ${\cal T}$. Rather, it is concentrated in an increasingly narrow interval of ${\cal O}$ conform to the decreasing width of the distribution. 
As to the certainty (Fig. \ref{fig4}c), it rather quickly converges upon increasing ${\cal T}$ to finite and rather big values in a wide interval of ${\cal O}$. This does not imply  that the distributions $P^{\pm}$ are different in this limit, since they become concentrated with divergent probability density, and the values of ${\cal O}$ with high certainty occur with exponentially low probability, yet the finite limit of $P_+-P_-$ is worth noting and deserves an explanation. 

We can qualitatively explain these features assuming that in this limit the probability distributions are the Gaussians with a shift that depends on the post-selection state and the variance $\sigma^2 = t_a/4 {\cal T}$, $P_{\pm} = g({\cal O}\pm s_{\pm}({\cal T}))$. In the limit of big ${\cal T}$ we expect the difference of the shifts to be proportional  
$({\cal T})^{-1}$, $s^{\pm} = \langle {\cal O}\rangle \pm S (t_a/{\cal T})$, $S \simeq 1$. This is because the effect of the post-selection is only felt during a time interval $\simeq \gamma^{-1}$ before the end of measurement, so that, at a fraction of the whole interval that is proportional to $ ({\cal T})^{-1}$. With this, at ${\cal O} \simeq \sigma$ the difference of the probabilities approaches a limit not depending on ${\cal T}$
\begin{equation}
P_+-P_- = \frac{S}{2} \frac{({\cal O} - \langle {\cal O}\rangle )}{\sigma \sqrt{2\pi}}\exp\left(-\frac{({\cal O} - \langle {\cal O}\rangle )^2}{2 \sigma^2}\right),
\end{equation}
The maximum difference of probabilities $|P_+-P_-|_{{\rm max}} \approx 1.9 S$ is thus achieved at ${\cal O} = \langle {\cal O}\rangle \pm\sigma$.

As to the certainty, it approaches an alternative limit at ${\cal O} \simeq 1 \gg \sigma$ that also does not depend on ${\cal T}$ at ${\cal T} \to \infty$
\begin{equation}
C({\cal O}) = \frac{P_+(\mathcal{O}) - P_-(\mathcal{O})}{P_+(\mathcal{O}) + P_-(\mathcal{O})} = {\rm tanh}\left( 4 S (\mathcal{O} - \langle {\cal O}\rangle) \right)
\end{equation}
As we see, the certainty reaches $\pm 1$ in the limit of large (exponentially improbable) $|{\cal O}| \gg 1$.

The numerical results presented are satisfactory fitted by above expressions with $S \approx 0.04$. However, the fits are not mathematically exact since, for the sake of simplicity, the shifts $s^{\pm}$ have been assumed not to depend on ${\cal O}$ while in general they do. 

Our results show that the difference of the conditioned distributions can be detected under realistic experimental circumstances.

Although the interference signature seem to disappear for rather short ${\cal T}$ in a realistic experimental regime, the actual measurements are done \cite{Huard} for  time intervals yet smaller than the time scale of qubit relaxation/decoherence. This correspond to the first several choices of short time intervals in Figures \ref{fig2}, \ref{fig3}, and \ref{fig4} where the interference is still visible.\\

\section{Numerical results: short time scales}
\label{sec:shortscale}
In the previous Section, we have considered the statistics at time-scales ${\cal T} \simeq \gamma^{-1},t_a$ extending the analytical results of Section \ref{sec:decscale}. In this Section, we will extend the analytical results of Section \ref{sec:suddenjump}.
We present numerical solutions for the probability distributions at a larger time-scale  $ {\cal T}\Omega \simeq 1$ of the Hamiltonian dynamics where the decoherence and relaxation does not play an important role. We also consider smaller ${\cal T}$ where  the {\it sudden jump} behavior is manifested, and yet smaller ${\cal T}$ where the decoherence becomes important again and the time-averaged output saturates to the value $\simeq \Omega/\gamma \gg 1$. We restrict ourselves to the experimental circumstances and 
use for the computation the Eq. \eqref{eq3exp} with the parameters specified in Section \ref{sec:decscale}.\\

\begin{figure*}[!ht]
\centering
\includegraphics[scale=0.5]{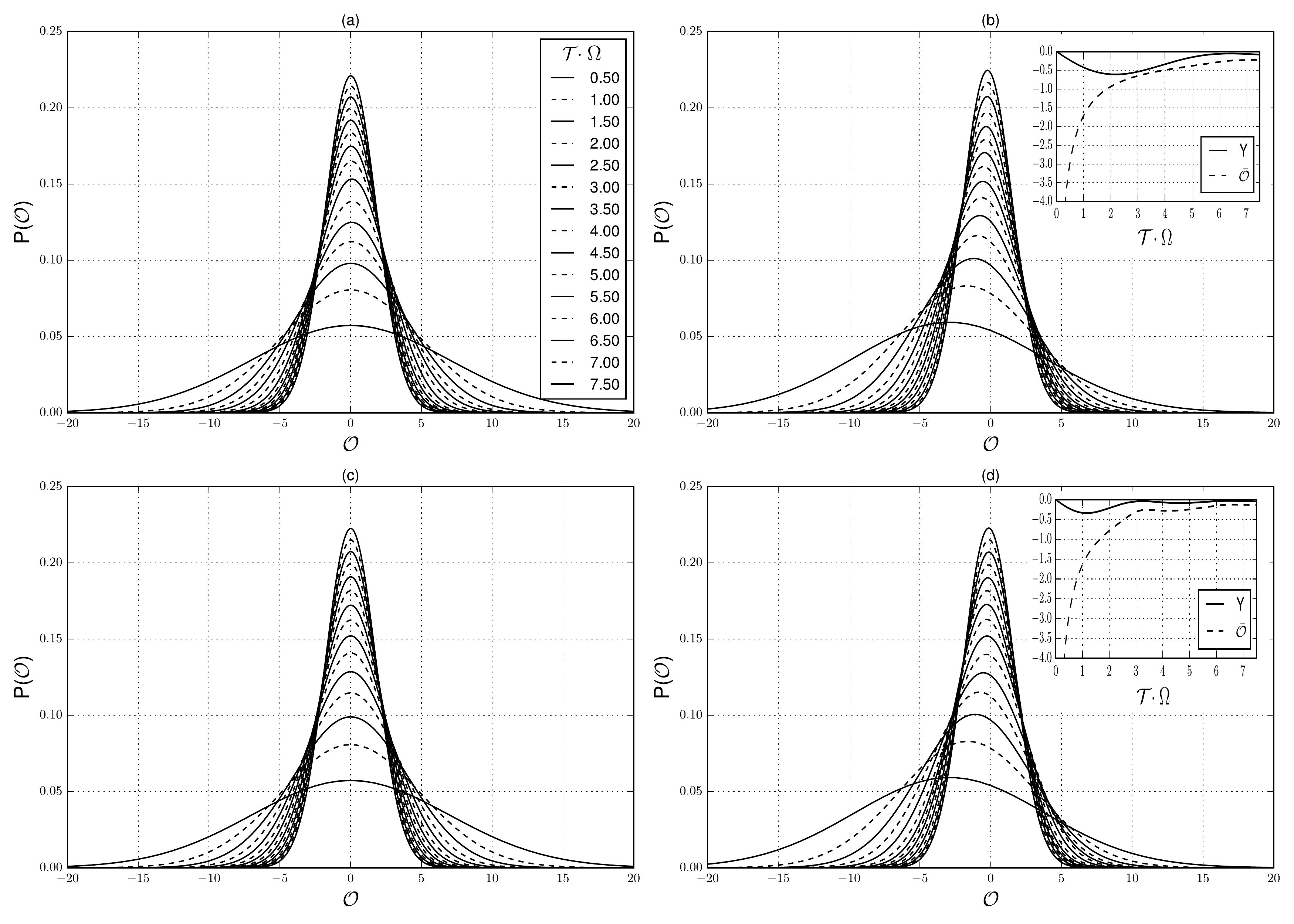}
\caption{The output distributions for the $\hat{\sigma}_y$ measurements for a series of ${\cal T}$ values at the scale of $\Omega^{-1}$ shown in the label of Figure (a). The qubit is initially prepared in the state $Z^+$. Left column: the distributions conditioned on $Z^+$. Right column: the distributions conditioned on $Z^-$.  Upper row: $\Delta=0$. Lower row: $\Delta = 1.7 \Omega$. The insets in the right column plots present the unconditioned average (solid curves) and the average of the distribution conditioned on $Z^-$. The distributions conditioned on $Z^+$ are symmetric with zero average. The values of the parameters correspond to \cite{Huard}. }
\label{fig5}
\end{figure*}

We will concentrate on the conditioned measurement statistics of the variable $\hat{\sigma}_y$, that anticommutes with the qubit Hamiltonian $\hat{H}_q=\frac{\hbar}{2}\Omega\hat{\sigma}_x+\frac{\hbar}{2}\Delta\hat{\sigma}_z$. The qubit is initially prepared in $Z^{+}$ state and post-selected in either $Z^{+}$ or $Z^{-}$.
In Fig. \ref{fig5}, the probability distributions of the integrated output ${\cal O}$ are presented.  The upper row  plots (Figs. (a) and (b)) are for zero detuning ($\Delta = 0$), while the lower row plots (Figs. (c) and (d)) show the corresponding distributions when at the detuning $\Delta\approx1.7\Omega$ that maximizes $\langle \sigma_x \rangle$ .\\
Left and right figures correspond to post-selection in $Z^{+}$ and $Z^{-}$, respectively.

For unconditioned distributions, the average output is given by
$Y({\cal T}) = \frac{1}{\cal T}\int_0^{{\cal T}}dt \langle \Psi(t)|\sigma_y| \Psi(t)\rangle$,
where $| \Psi(t)\rangle$ is obtained from $Z^+$ by Hamiltonian evolution. The function $Y({\cal T})$ is plotted in the insets of the right plots with a solid curve. 
We would expect the distributions to be shifted with respect to the origin by a value ${\cal O} \simeq 1$. This shift would be clearly seen in the plots since the width of the distribution $\simeq \sqrt{t_a\simeq {\cal T}} \simeq \sqrt{t_a \Omega}$ is not very big at experimental values of $\Omega t_a \approx 200$. However, the plots on the left are perfectly centered at the origin at any ${\cal T}$. Indeed, the zero average of the distributions conditioned at $Z^+$ can be proven analytically in the limit of Hamiltonian dynamics. 
The averages of the distributions conditioned at $Z^-$ (given by dashed curves in the insets of the plots) increase at small ${\cal T}$  as ${\cal T}^{-1}$, in agreement with Eq. \ref{eq:suddenjump}. The ratio of this average to conditioned average is just the inverse probability to be found in $Z-$, $p_-({\cal T }) = \sin^2(\sqrt{\Omega^2+\Delta^2}){\cal T}/2) /(1+(\Delta/\Omega)^2)$, $p_- \propto {\cal T}^2$ at small ${\cal T}$.

These averages are visually manifested as the shifts of the distributions that are largely Gaussian. We do not see anything resembling a gap in the distribution predicted for an ideal detector (Fig. \ref{fig:shortscales}). This is explained by relatively low detection efficiency (c.f. Eq. \ref{eq:distshorttimes}).

\begin{figure*}[!ht]
\centering
\includegraphics[scale=0.6]{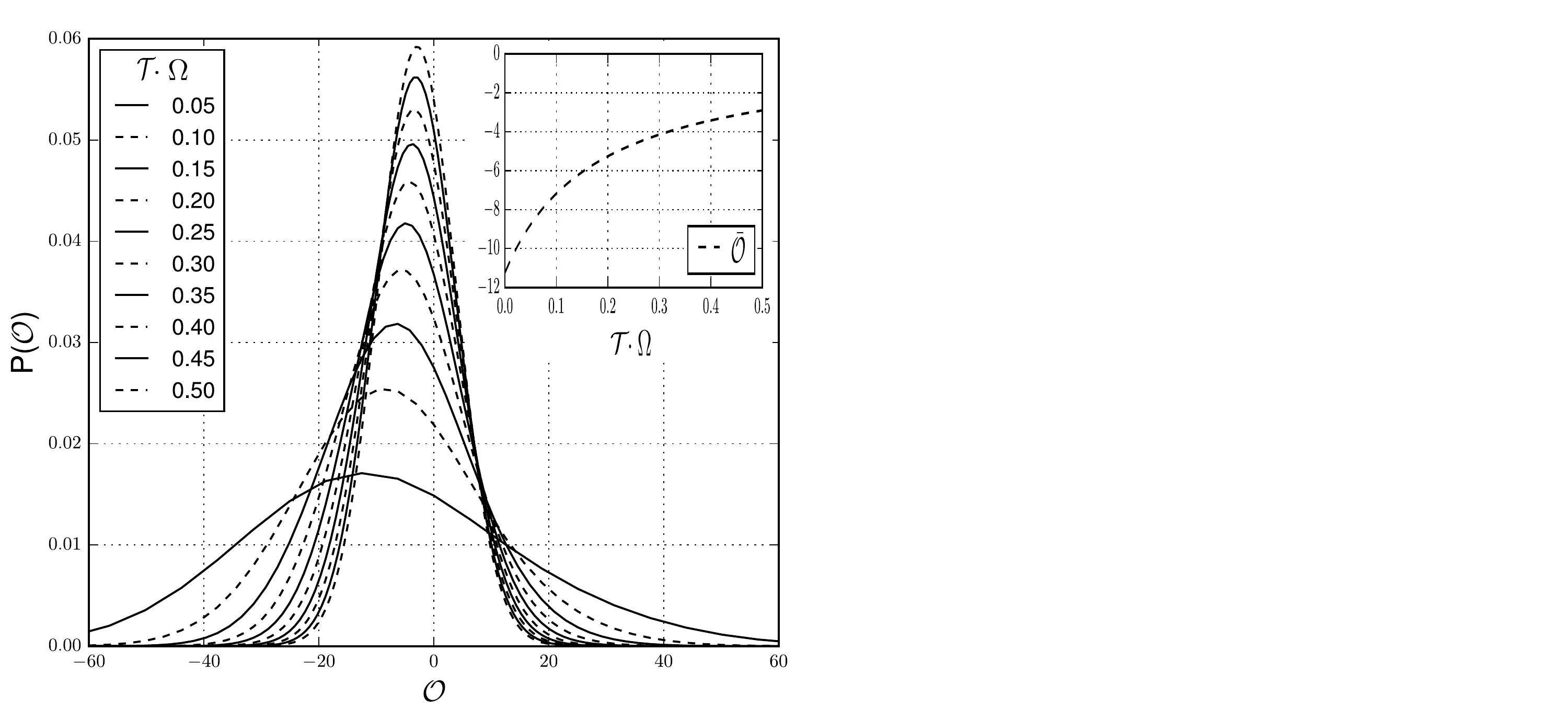}
\caption{The probability distributions of outcomes of $\hat{\sigma}_y$ measurement for different ${\cal T} \ll \Omega^{-1}$. The values of the parameters correspond to \cite{Huard}.
The qubit is prepared in $Z^{+}$ and post-selected in an orthogonal state $Z^{-}$. The average output signal $\bar{\cal O}$ is shown in the inset. It exhibits anomalously large values and saturation in the limit ${\cal T} \to 0$.}
\label{fig6}
\end{figure*}

In a separate Fig. \ref{fig6} we present the distributions conditioned on $Z^{-}$ at yet smaller time-scales of the order of the sudden jump duration (see Eq. \ref{eq:shorterscales}). In this regime, we see the saturation of the average $\bar{\cal O}$ at a value close to $-11$ in the limit ${\cal T}\to 0$. This gives the upper limit of anomalously big averages under experimental conditions of \cite{Huard}.
The distributions can be well approximated by shifted Gaussians, smaller ${\cal T}$ corresponding to wider distributions. 

\section{Conclusion}
\label{sec:conclusion}
Recent experimental progress has enabled the measurements in course of the conditioned quantum evolution. The average signals have been experimentally studied in \cite{Huard, DiCarlo, SiddiqiMolmer}. The technical level of these experiments permits the characterization of  the complete statistics of the measurement outputs.\\
In this work, we have developed a proper theoretical formalism based on full counting statistics approach \cite{NazWei, NazKin} to describe and evaluate these statistics.
We illustrate it with several examples and prove that the interesting features  in statistics can be seen in experimentally relevant regimes (Fig. \ref{fig3} and \ref{fig6}), for both short and relatively long measurement time intervals.

We reveal and investigate analytically two signatures of the conditioned statistics that are related to quantum interference effects. First is the {\it half-quantized} measurement values. We demonstrate that the conditioned distribution function may display peculiarities --- that are either peaks or dips --- at {\it half-sums} of the quantized values.

Second signature pertains the case of zero overlap between initial and final state and time intervals that are so short as the wave function of the system does not significantly change by either Hamiltonian or dissipative dynamics.

We reveal unexpectedly large values of the time-integrated output cumulants for such short intervals, that we term {\it sudden jump}. We show that the account for decoherence leads to a finite duration of the jump at ultra-short time-scale $\gamma/(\Omega^2)$ and saturation of the anomalous eigenvalues at $\Omega/\gamma$, $\Omega$ and $\gamma$ being the frequency scales of the Hamiltonian and dissipative dynamics, respectively.

Actually, we have shown with our results that one can have very detailed theoretical predictions of CWLM distributions that can account for every detail of the experiment. This enables investigation and characterization of quantum effects even if the choice of parameters is far from the optimal one and these effects are small.
We emphasize once again that the interference signature in the distributions that we predict in this Article can be seen in realistic experimental regimes and hope the effects can be experimentally observed soon. The efficient recording of time traces for a weak continuous monitoring of one, or several, qubit variables, is a key ingredient for accessing these statistics. It has been achieved in several articles and applied for observation of  single quantum "trajectories" or real time feedback.\cite{SiddiqiFeedback} High fidelity preparation and post-selection of the qubit is also required for experiments with conditioned evolution, yet this is a general requirement in most qubit experiments. We thus believe that it is possible to extract the interesting statistics from the existing records.\\

This work was supported by the Netherlands Organization for Scientific Research (NWO/OCW), as part of the Frontiers of Nanoscience program.

\bibliography{authorbibtex}
\end{document}